\documentclass[11pt]{article}

\usepackage[utf8]{inputenc}
\usepackage{amsmath, amssymb, amsfonts, graphicx}
\usepackage[colorinlistoftodos]{todonotes}
\usepackage[colorlinks=true, allcolors=blue]{hyperref}
\usepackage{listings, color}
\usepackage[normalem]{ulem}

\usepackage{subfig}
\usepackage{enumitem}
\usepackage{float}
\usepackage{textcomp}
\usepackage{authblk}
\usepackage[numbers,super]{natbib}
\usepackage{hyperref}
\usepackage{fullpage}
\usepackage{multirow}
\usepackage{url}
\usepackage{arydshln}
\usepackage{wrapfig}

\begin{document}

\title{Lossy compression of statistical data using quantum annealer}

\author[1\footnote{boram@lanl.gov}]{Boram Yoon}
\affil[1]{CCS-7, Computer, Computational and Statistical Sciences Division, \protect\\Los Alamos National Laboratory,Los Alamos, NM 87545, USA}

\author[2\footnote{nga.nguyen@lanl.gov}]{Nga T.T. Nguyen}
\affil[2]{CCS-3, Computer, Computational and Statistical Sciences Division, \protect\\Los Alamos National Laboratory,Los Alamos, NM 87545, USA}

\author[3,4,5\footnote{chiacheng.chang@riken.jp}]{Chia Cheng Chang}
\affil[3]{RIKEN iTHEMS, Wako, Saitama 351-0198, Japan}
\affil[4]{Department of Physics, University of California, Berkeley, California 94720, USA}
\affil[5]{Nuclear Science Division, Lawrence Berkeley National Laboratory, \protect\\Berkeley, California 94720, USA}

\author[4\footnote{ermalrrapaj@berkeley.edu}]{Ermal Rrapaj}

\date{}

\maketitle

\begin{abstract}

We present a new lossy compression algorithm for statistical floating-point data through a representation learning with binary variables.
The algorithm finds a set of basis vectors and their binary coefficients that precisely reconstruct the original data. The optimization for the basis vectors is performed classically, while binary coefficients are retrieved through both simulated and quantum annealing for comparison.
A bias correction procedure is also presented to estimate and eliminate the error and bias introduced from the inexact reconstruction of the lossy compression for statistical data analyses.
The compression algorithm is demonstrated on two different datasets of lattice quantum chromodynamics simulations. The results obtained using simulated annealing show 3.5 times better compression performance than the algorithms based on a neural-network autoencoder and principal component analysis. Calculations using quantum annealing also show promising results, but performance is limited by the integrated control error of the quantum processing unit, which yields large uncertainties in the biases and coupling parameters. Hardware comparison is further studied between the previous generation D-Wave 2000Q and the current D-Wave Advantage system. Our study shows that the Advantage system is more likely to obtain low-energy solutions for the problems than the 2000Q.
\end{abstract}

\section{Introduction}\label{sec:intro}
Today's scientific computing and experiments often produce petabytes of floating-point data that need to be stored for post-processing or transferred to different computing centers. For example, modern lattice quantum chromodynamics (QCD) simulations targeting accurate precision generate $O(\textrm{PB})$ of data~\cite{Park:2021ypf, He:2021yvm} and store the data on storage systems for long-term analysis. In many applications, only a few significant figures of the stored data are required for the analysis, so lossy data compression algorithms are considered as viable approaches to reducing the data storage requirement and increasing the effective bandwidth for data movement.

Various lossy data compression algorithms recently proposed for floating-point arrays of scientific data include ISABELA~\cite{osti_1564924}, ZFP~\cite{6876024}, SZ~\cite{7516069, Tao_2017, 10.1145/3410463.3414624}, and NUMARCK~\cite{7013047}. ISABELA provides in-situ compression based on interpolation using B-splines~\cite{135914} after sorting multidimensional scientific data. ZFP uses block transformation for decorrelation and bit-plane encoding for a fixed-rate lossy compression. SZ is an error-bounded lossy compression algorithm based on fitting and predicting the successive data points. NUMARCK achieves the data compression by approximating the temporal changes using the K-means data clustering algorithm~\cite{jin2006}.

For statistical data, it is possible to detect the correlation pattern of the data components using machine learning techniques and exploit the learned correlation for efficient lossy data compression. An example is the approaches based on the autoencoder~\cite{PhysRevFluids.5.114602,WANG2020105761,Romero_2017}. Unsupervised machine learning techniques allow us to find efficient codings of the input data. They work as data compression algorithms since the coding is typically a lower-dimensional representation of the original data. The compression performance of the representation learning can be maximized by restricting the codes to be binary variables so that each code can be stored in a bit. However, an encoder with binary codes generally involves binary optimization for finding the optimal codes, which is an NP-hard problem.

Such binary optimization can be solved by Ising solvers such as
D-Wave quantum annealers. The quantum processor of the D-Wave systems finds low-lying energy states of the target Ising Hamiltonian starting from the transverse field Hamiltonian through quantum annealing~\cite{dwave_system}. In general, quantum annealing will require an exponentially large number of samples in order to recover the most optimal solution to large binary optimization problems. However, the annealing time for each sample takes only $O(10)$ microseconds. Furthermore, we show in this work that low-energy solutions, which can be obtained from only a small number of samples, are sufficient for our proposed sparse coding compression algorithm.

In this paper, we propose a new data compression algorithm based on a representation learning. For a maximum compression, we use binary codes, and formulate the problem in a quadratic unconstrained binary optimization (QUBO) form so that it can be solved on Ising solvers such as the D-Wave quantum annealer, as described in Section~\ref{sec:method}. As a result, the algorithm guarantees the compression ratio while optimizing for the smallest loss. We also present a bias correction procedure that removes the bias and estimate the errors due to the inexact reconstruction from the lossy-compressed statistical data. In Section~\ref{sec:expt}, the proposed compression algorithm is demonstrated for two different lattice QCD datasets using simulated annealing, and D-Wave's 2000Q and the Advantage System.

\section{Method}\label{sec:method}
\subsection{Data Compression}\label{sec:alg}
The goal of this algorithm is to find a matrix $\boldsymbol{\phi}\in \mathbb{R}^{D\times N_q}$ and binary coefficients $\boldsymbol{a}^{(k)}\in \{0,1\}^{N_q}$ that precisely reconstruct the input vectors $\mathbf{X}^{(k)}\in \mathbb{R}^{D}$ such that $\mathbf{X}^{(k)} \approx \boldsymbol{\phi} \boldsymbol{a}^{(k)}$ for all data index $k=1,2,3,\ldots, N$. The procedure defines a mapping from $\mathbf{X}$-space to $\boldsymbol{a}$-space:
\begin{align}
\left\{\mathbf{X}^{(k)}\vert\mathbf{X}^{(k)}\in \mathbb{R}^{D}, k=1,2,\ldots,N\right\}
\longrightarrow 
\left(\left\{\boldsymbol{a}^{(k)} \vert \boldsymbol{a}^{(k)}\in \{0,1\}^{N_q}, k=1,2,\ldots,N\right\}, \boldsymbol{\phi}\in \mathbb{R}^{D\times N_q}\right)\,.
\end{align}
Here the coefficients in the $\boldsymbol{a}$-space are restricted to binary variables so that it can be stored in a single bit. Additionally, we
restrict $N_q \ll N$ so that the memory usage of $\boldsymbol{\phi}$ is comparatively small to the uncompressed data, which for high-statistics datasets where compression is necessary is of $N\gtrsim\mathrm{O}(10^4)$. 
As a result, the data in $\boldsymbol{a}$-space uses less memory space than those in $\mathbf{X}$-space, and results in data compression. 

One possible solution of the mapping ($\{\boldsymbol{a}^{(k)}\}$ and $\boldsymbol{\phi}$) can be obtained by minimizing the mean square error of the reconstruction as following:
\begin{align}
	\min\limits_{ \boldsymbol{\phi} }\sum_{k}\min\limits_{ \boldsymbol{a}^{(k)} } \left[  \,   \sum_{i=1}^D 
	\left( X^{(k)}_i - \big[\boldsymbol{\phi} \boldsymbol{a}^{(k)}\big]_i \right)^2 \, \right]\,.
	\label{eq:H_SC}
\end{align}
When the underlying data exhibits heteroskedasticity, a weight factor of inverse variance, $1/\sigma_{X_i}^2$, needs to be multiplied to each term of the least-squares loss function to avoid the algorithm focusing on the reconstruction of the large-variance components of the input vector and to make the reconstruction error uniform. The same effect can be achieved by standardizing the input data $\mathbf{X}$ in the data preparation.
The resulting optimization problem is mapped to a QUBO
\begin{equation}
	H(\boldsymbol{h}, J, \boldsymbol{s}) = \sum_i^{N_q} {h_i s_i} + \sum_{i<j}^{N_q} {J_{ij}  s_i s_j }\,,
	\label{eq:QUBO}
\end{equation}
through the transformation given below:
\begin{align}
J = 2\boldsymbol{\phi}^T\boldsymbol{\phi}, \qquad {h}_i = -2\left[\boldsymbol{\phi}^T \mathbf{X}\right]_i + \left[\boldsymbol{\phi}^T\boldsymbol{\phi}\right]_i, \qquad \boldsymbol{s} = 2\boldsymbol{a} -1\,.
\label{eq:QUBO-transf}
\end{align}
Note that structure of the transformation is similar to the one used in sparse coding~\cite{Nguyen:2016,8123653,Nguyen2018ImageCU}, but the compression algorithm does not require the constraints $\left[\boldsymbol{\phi}^T\boldsymbol{\phi}\right]_i=1$, placed in the sparse coding.


After obtaining the solution of $\boldsymbol{a}^{(k)}$ for a given $\boldsymbol{\phi}$,
we update $\boldsymbol{\phi}$ using stochastic gradient decent on a classical computer. The optimizations for $\boldsymbol{a}^{(k)}$ and $\boldsymbol{\phi}$ are iterated until they reach to a stationary solution. The procedure can be summarized as following.
\begingroup
\renewcommand\labelenumi{(\theenumi)}
\begin{enumerate}
  \item Initialize $\{\boldsymbol{a}^{(k)}\}$ and $\boldsymbol{\phi}$ with random numbers or initial guesses.
  \item Take a random mini-batch of size $N_b$ from the $N$ samples of $\mathbf{X}^{(k)}$.
  \item Within the mini-batch, fix $\boldsymbol{\phi}$ and find $\{\boldsymbol{a}^{(k)}\}$ that minimizes Eq.~\eqref{eq:H_SC}.
  \item Within the mini-batch, fix $\{\boldsymbol{a}^{(k)}\}$ and update $\boldsymbol{\phi}$ towards the optimum solution of Eq.~\eqref{eq:H_SC} with a learning rate $\eta$.
  \item Repeat (2)--(4) until it reaches the minimum reconstruction error.
\end{enumerate}
\endgroup
\noindent
Here the mini-batch size $N_b$ and the learning rate $\eta$ control the convergence of the algorithm.

\subsection{Bias Correction}\label{sec:BC}
In many scientific applications, such as the Monte Carlo simulations, our major concern is the expectation value of a function of the statistical variables $\langle f(\mathbf{X})\rangle$. With the samples $\mathbf{X}^{(k)}$, the expectation value is usually estimated by a simple average over $k$. When using the compressed data in $\boldsymbol{a}$-space, however, the lossy-compression introduces reconstruction error $\mathbf{X}^{(k)} \neq \boldsymbol{\phi} \boldsymbol{a}^{(k)} \equiv \tilde{\mathbf{X}}^{(k)}$. As a result, a simple average $\frac1N\sum_{k=1}^N f(\tilde{\mathbf{X}}^{(k)})$ as an estimator of $\langle f(\mathbf{X})\rangle$ is biased.

An unbiased estimator $\bar{O}^{\textrm{BC}}$ can be defined by using a small portion of the original data $\mathbf{X}^{(k)}$:
\begin{align}
  \bar{O}^{\textrm{BC}} = \frac1N\sum_{k=1}^N f(\tilde{\mathbf{X}}^{(k)})
  + \frac{1}{N_{\textrm{bc}}}\sum_{k=1}^{N_{\textrm{bc}}} \left( f(\mathbf{X}^{(k)}) - f(\tilde{\mathbf{X}}^{(k)})\right)\,.
  \label{eq:unbiased}
\end{align}
Here the first term on the right hand side is a sloppy estimator of $\langle f(\mathbf{X})\rangle$, and the second term is a bias correction term that makes the estimator satisfy $\langle \bar{O}^{\textrm{BC}} \rangle = \langle f(\mathbf{X})\rangle$.  Note that in the second term, we use the first $N_{\textrm{bc}}$ samples out of total $N$ samples as a bias correction dataset, assuming the data samples are independent and identically distributed. Depending on the data characteristics, however, one could take the maximally separated or randomly chosen $N_{\textrm{bc}}$ samples for the bias correction dataset.

In addition to the $\{\boldsymbol{a}^{(k)}\}$ and $\boldsymbol{\phi}$, for a bias correction in the reconstruction, one needs to store the $N_{\textrm{bc}}$ samples of the original data $\{\mathbf{X}^{(k)}\vert k=1,2,\ldots,N_{\textrm{bc}}\}$. As explained in Section~\ref{sec:secQ}, the statistical error of $\bar{O}^{\textrm{BC}}$ induced by the bias correction term depends on $N_{\textrm{bc}}$ and the correlation between $f(\mathbf{X}^{(k)})$ and $f(\tilde{\mathbf{X}}^{(k)})$. For a good compression, which yields high correlation between correlation between $f(\mathbf{X}^{(k)})$ and $f(\tilde{\mathbf{X}}^{(k)})$, the bias $f(\mathbf{X}) - f(\tilde{\mathbf{X}}^{(k)})$ can be estimated precisely from a small number of samples, so one can take $N_{\textrm{bc}} \ll N$. A similar structure of bias correction has been demonstrated in the machine learning regressions on statistical data~\cite{Yoon:2018krb, Zhang:2019qiq}.

In the calculation of the statistical error of $\bar{O}^{\textrm{BC}}$, the correlation between the sloppy estimator and the bias correction term should be taken into account. One approach to make the procedure simple is binning the data so that each bin has a certain number of bias correction data samples and the data in different bins are uncorrelated with each other. Assuming that the number of bins $N_{\textrm{bin}}$ divides $N$ and $N_{\textrm{bc}}$, Eq.~\eqref{eq:unbiased} can be rewritten as
\begin{align}
  \bar{O}^{\textrm{BC}} &= \frac{1}{N_{\textrm{bin}}} \sum_{i=1}^{N_{\textrm{bin}}} \left[
  \frac{1}{M}\sum_{k=iM+1}^{(i+1)M} f(\tilde{\mathbf{X}}^{(k)})
  + \frac{1}{M_{\textrm{bc}}}\sum_{k=iM+1}^{iM+M_{\textrm{\textrm{bc}}}} \left( f(\mathbf{X}^{(k)}) - f(\tilde{\mathbf{X}}^{(k)})\right)
  \right] \\
  &\equiv \frac{1}{N_{\textrm{bin}}} \sum_{i=1}^{N_{\textrm{bin}}} \bar{O}^{\textrm{BC},b}_i\,,
\end{align}
where $M = N/N_{\textrm{bin}}$ and $M_{\textrm{bc}} =  N_{\textrm{bc}}/N_{\textrm{bin}}$. In this rearrangement, the statistical error of $\bar{O}^{\textrm{BC}}$ can be calculated by $\sigma_{\bar{O}^{\textrm{BC}}} = \sigma_{\bar{O}^{\textrm{BC},b}}/\sqrt{N_{\textrm{bin}}}$. Again, note that the first $M_{\textrm{bc}}$ samples in each bin are used for the bias correction, but one can take maximally separated or randomly chosen samples for the bias correction dataset, depending on the characteristics of the data.

\subsection{Quality Indicator for Lossy-Compression}\label{sec:secQ}
To measure the quality of lossy-compression on statistical data, we define the $Q^2$ as
\begin{align}
  Q^2 \equiv \frac1D \sum_{i=1}^D \frac{\sigma_{X_i-\tilde{X}_i}^2}{\sigma_{X_i}^2} \,,
  \label{eq:Q-value}
\end{align}
where $\sigma^2_{X_i-\tilde{X}_i}$ is the variance of ${X_i-\tilde{X}_i}$. This parameter is an indicator of the statistical error increase due to the lossy-compression after the bias correction as following. Consider a simple bias-corrected average of independent observables
\begin{align}
  \bar{\mathbf{X}}^{\textrm{BC}} = \frac1N\sum_{k=1}^N \tilde{\mathbf{X}}^{(k)}
  + \frac{1}{N_{\textrm{bc}}}\sum_{k=1}^{N_{\textrm{bc}}} \left( \mathbf{X}^{(k)} - \tilde{\mathbf{X}}^{(k)}\right)\,.
  \label{eq:unbiased_avg}
\end{align}
The variance of the $i$-th component of $\bar{\mathbf{X}}$ can be approximated as
\begin{align}
 \sigma^2_{\bar{X}_i^{\textrm{BC}}} 
  &\approx \frac{1}{N} \sigma^2_{\tilde{X}_i} + \frac{1}{N_{\textrm{bc}}}\sigma^2_{X_i-\tilde{X}_i} \\
 &\approx \frac{\sigma^2_{X_i}}{N} 
   \left( 1 + \frac{N}{N_{\textrm{bc}}} \frac{\sigma^2_{X_i-\tilde{X}_i}}{\sigma^2_{X_i}} \right)\,,\label{eq:var_approx}
\end{align}
where the first approximation assumes a small correlation between the two terms in Eq.~\eqref{eq:unbiased_avg}, and the second approximation assumes a good lossy-compression that gives $\sigma^2_{\tilde{X}_i} \approx \sigma^2_{X_i}$. 
Assuming a small reconstruction error satisfying $\sigma^2_{X_i-\tilde{X}_i} /\sigma^2_{X_i} \ll N/N_{\textrm{bc}}$, the expected statistical error increase due to the bias correction can be estimated as
\begin{align}
 \frac{\sigma_{\bar{X}_i^{\textrm{BC}}}}
 {\sigma_{\bar{X}_i}}
  \approx 
    1 + \alpha\frac{N}{2N_{\textrm{bc}}} \frac{\sigma^2_{X_i-\tilde{X}_i}}{\sigma^2_{X_i}}\,,\label{eq:err_inc}
\end{align}
where $\alpha=1$ and $\sigma^2_{\bar{X}_i} = \sigma^2_{X_i}/N$. It shows that the increase of the statistical error compared to that of the original data is proportional to ratio of the number of bias correction data, $N/N_{\textrm{bc}}$, and the normalized variance of the reconstruction error, $\sigma^2_{X_i-\tilde{X}_i} /\sigma^2_{X_i}$. Hence, we define the quality of the compression by taking an average of $\sigma^2_{X_i-\tilde{X}_i} /\sigma^2_{X_i}$ over the all vector elements as given in Eq.~\eqref{eq:Q-value}.

Note that, when data have autocorrelation, the bias correction dataset can be chosen such that they have smaller autocorrelation than the original data by taking a wide separation in the trajectory direction of the autocorrelation. It makes the bias correction more efficient, suppresses the statistical error increase, and yields $\alpha<1$.

\subsection{Boosting}\label{sec:boosting}
In practice, the binary optimization in Eq.~\eqref{eq:H_SC} is difficult to solve for a large $N_q$. Although a quantum annealer is employed to solve the optimization problem, the maximum number of fully-connected qubits is limited to $\mathcal{O}(100)$ for the current quantum processors. However, the problem can be decomposed into a linear combination of smaller $N_q$ by applying the idea of Boosting~\cite{k-thb-88, Freund99ashort}.

Assume that we have a matrix $\boldsymbol{\phi}_1$ and vectors $\{\boldsymbol{a}_1^{(k)}\}$ of $N_{q1}$ binary elements that approximately reconstruct the input vectors $\mathbf{X}^{(k)} \approx \boldsymbol{\phi}_1 \boldsymbol{a}^{(k)}_1$. We can find another set of solutions of $\boldsymbol{\phi}_2$ and $\{\boldsymbol{a}_2^{(k)}\}$ of $N_{q2}$ binary elements reconstructing the reconstruction error of $\boldsymbol{\phi}_1 \boldsymbol{a}^{(k)}_1$ by taking $\mathbf{X}^{(k)} - \boldsymbol{\phi}_1 \boldsymbol{a}^{(k)}_1$ as new input vectors. By combining the two sets of solutions, we can build a precise reconstruction of $\mathbf{X}^{(k)}$ as
\begin{align}
  \mathbf{X}^{(k)} \approx \boldsymbol{\phi}_1 \boldsymbol{a}^{(k)}_1 + \boldsymbol{\phi}_2 \boldsymbol{a}^{(k)}_2
  = \begin{pmatrix}
\boldsymbol{\phi}_1 & 0 \\
0 & \boldsymbol{\phi}_2
\end{pmatrix}
\begin{pmatrix}
\boldsymbol{a}^{(k)}_1 \\
\boldsymbol{a}^{(k)}_2
\end{pmatrix}
  \,,
\end{align}
where the total number of binary coefficients representing an input vector $\mathbf{X}^{(k)}$ is $N_{q1}+N_{q2}$. This procedure can be repeated to an arbitrary number of sets of solutions. For an ideal binary optimizer, decomposing the problem into a smaller number of binary elements makes the solution worse than the full solution because the decomposition ignores the correlation between the different sets, that potentially reduces the reconstruction error. For realistic binary optimizers, however, boosting can provide a better solution.

\section{Numerical Experiments}\label{sec:expt}
\subsection{Test Data}
\begin{figure}
    \centering
    \includegraphics[width=0.45\textwidth]{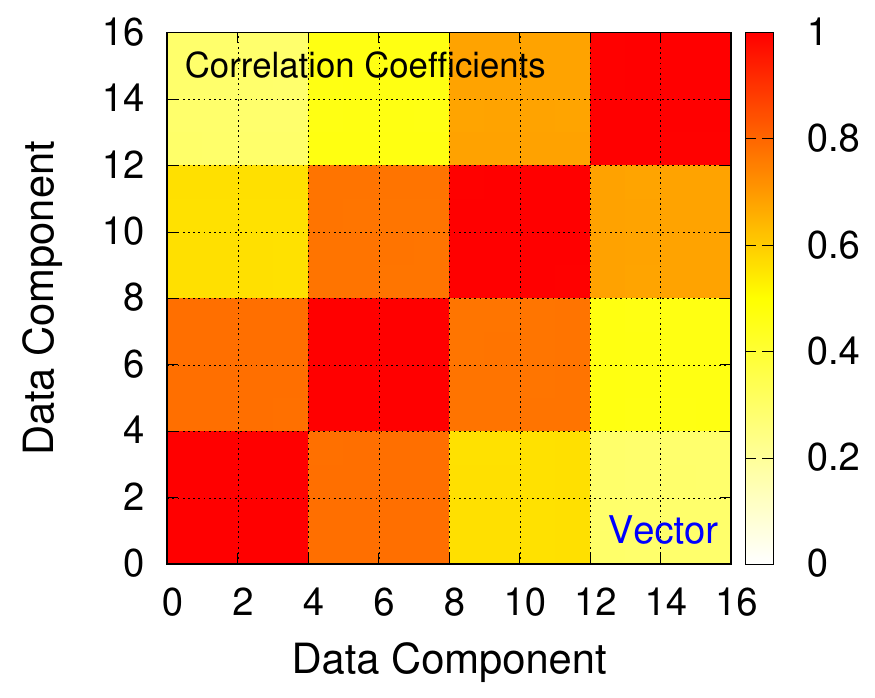} \qquad
    \includegraphics[width=0.45\textwidth]{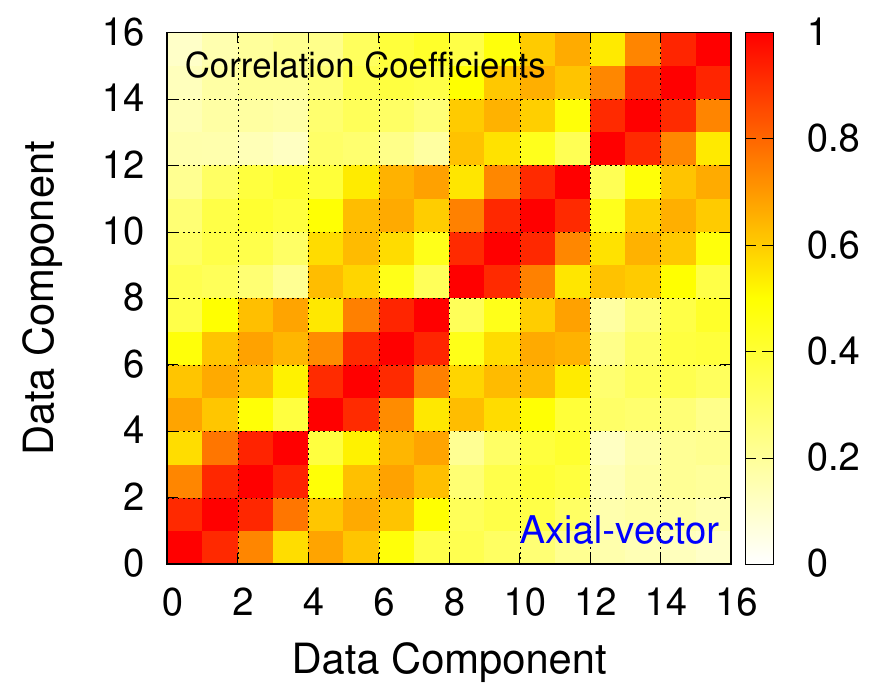}
    \caption{Correlation pattern of the 16 components of the vector (left) and the axial-vector (right) data. Red indicates the high correlation (correlation coefficient = 1), and white indicates no correlation.}
    \label{fig:data}
\end{figure}

In this study, we use the Monte Carlo simulation data of lattice QCD, a theory of quarks and gluons, and their interactions. The lattice QCD simulations produce large amounts of data that need to be stored for analysis, but the data are correlated with each other, so a data compression algorithm exploiting the correlation can obtain a better compression ability.
Among various lattice QCD observables, in this study, we use the three-point correlation function data of nucleon \emph{vector} and \emph{axial-vector (axial)} charges, which describe the response of a nucleon to particles such as the neutrino. We shape the data into 10 independent sets of 3200 vectors with 16 components, so each dataset has $N=3200$ and $D=16$. As illustrated in Figure~\ref{fig:data}, there are strong correlations between the 16 components. Since the \emph{vector} data show a stronger correlation than the \emph{axial-vector} data, we expect the proposed algorithm to give a better compression (smaller $Q^2$) for the \emph{vector} data than the \emph{axial-vector} data. We standardize the data as a pre-processing step to obtain a homogeneous reconstruction error on all 16 components.

\subsection{Experiments with Simulated Annealing}\label{sec:exp_sim}
First, we carry out the demonstration of the proposed compression algorithm using the D-Wave's simulated annealing sampler, implemented in the D-Wave's Ocean library~\cite{dwave_ocean}, on classical computers. In the simulated annealing, we take the minimum energy solution from the 150 runs (\emph{num\_reads}=150), while all other parameters are set to their default values. As described in Section~\ref{sec:perf_sim}, we find that the simulated annealing with \emph{num\_reads=150} gives close to the exact solution up to around $N_q=20$, and the quality of the solution deteriorates as $N_q$ is increased.

To find the solutions $\{\boldsymbol{a}^{(k)}\}$ and $\boldsymbol{\phi}$ of the optimization problem in Eq.~\eqref{eq:H_SC}, we iterate the optimization in $\boldsymbol{\phi}$ and $\{\boldsymbol{a}^{(k)}\}$ as described in Section~\ref{sec:alg}. In this study, $\boldsymbol{\phi}$ is updated as follows. After obtaining the solution $\tilde{\boldsymbol{\phi}}$ that minimizes the reconstruction error of the mini-batch using the L-BFGS-B algorithm~\cite{Zhu94l-bfgs-b-}, we update $\boldsymbol{\phi}$ as
\begin{equation}
  \boldsymbol{\phi} \leftarrow \boldsymbol{\phi} + \eta(\tilde{\boldsymbol{\phi}}-\boldsymbol{\phi})\,.
\end{equation}
Here the learning rate $\eta$ is continuously decreased from the initial value $\eta_0$ as the number of training epochs ($n_{\textrm{epoch}}$) is increased, following
$\eta = \eta_0 \times 0.8^{n_{\textrm{epoch}}}$. For the batch size and initial learning rate, we use $N_b=50$ and $\eta_0=0.9$ as we find that those give the best or close to the best results after exploring a grid of $N_b$ and $\eta_0$. The final results are obtained with 30 epochs of training steps.

To compare with the compression performance of the proposed algorithm, we study the conventional data compression algorithms using principal component analysis (PCA) and neural-network-based autoencoder. PCA finds orthogonal directions that maximize the variance as principal components. By saving only the coefficients of the first few principal components, the PCA works as a lossy data compression algorithm. We reconstructed the data from the first $N_z$ principal components to obtain the data compression. Autoencoder also provides data compression by constraining the number of codes ($N_z$) to a small number~\cite{theis2017lossy,8470336}. We used a fully connected neural-network  encoder and decoder with three hidden layers of $(D, 128, 64, 32, N_z)$ and $(N_z, 32, 64, 128, D)$ with rectified linear unit (ReLU) activation functions. For the training, we use the Adam optimizer~\cite{kingma2017adam} implemented in the PyTorch python library~\cite{NEURIPS2019_9015} with the learning range of 0.01 and the batch size of 3200, which are the optimal hyperparameters determined from a grid search. After 5000 epochs of training, we continue the training until we reach a better reconstruction error than the best reconstruction error we have obtained in the first 5000 epochs and stop the training.

\begin{figure}
    \centering
    \includegraphics[width=0.45\textwidth]{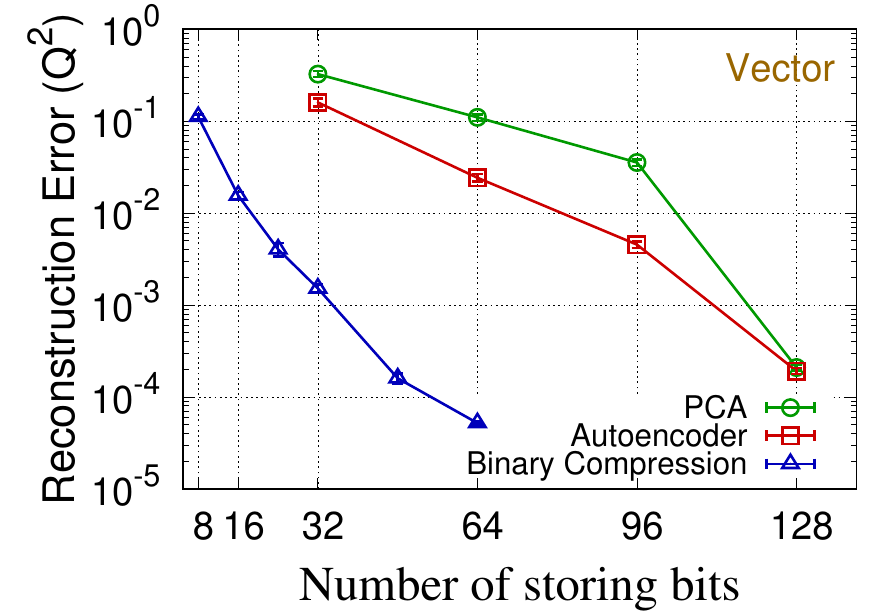}\qquad
    \includegraphics[width=0.45\textwidth]{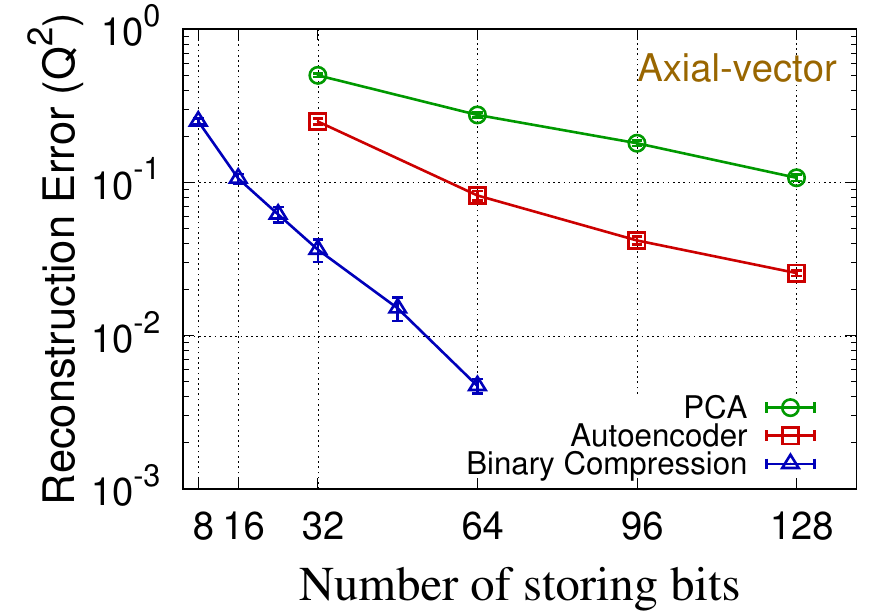}
    \caption{$Q^2$, defined in Eq.~\protect\eqref{eq:Q-value} for different number of storing bits. For the principal component analysis (PCA) and autoencoder (AE) approaches, the number of storing bits is calculated by $32\times N_z$, assuming single-precision floating-point numbers. For binary compression algorithm of $N_q >= 48$, we use the boosting approach with $N_{q1} = N_{q2} = N_q/2$.}
    \label{fig:Q-simann}
\end{figure}

The results are summarized in Table~\ref{tab:Q-simann} and Figure~\ref{fig:Q-simann}. The results show that the boosting approach gives better results than the full calculation for $N_q>32$, where the simulated annealing fails in finding the close-to-ground solution. The comparison between different algorithms shows that the autoencoder outperforms the PCA, and the proposed binary compression outperforms the autoencoder. The compression quality ($Q^2$) of the autoencoder with $N_z$ number of codes can be obtained using the proposed binary compression algorithm with the number of bits around $N_q\approx 9N_z$. Considering single-precision floating-point numbers, which usually occupying 32 bits for a number, the proposed algorithm provides the same quality of compression as the autoencoder approach using about 3.5 times smaller memory space. 

\begin{table}[htb]
    \centering
    \small
    \begin{tabular}{c|ll}
\hline\hline
$N_q$      & Vector  & Axial \\
($N_{q1}+N_{q2}$) & $Q^2$(BC) & $Q^2$(BC) \\\hline
 8    & 0.1115(65)   &  0.251(11)  \\\hdashline
16    & 0.0156(14)   &  0.1062(74) \\
8+8   & 0.0276(42)   &  0.1247(64) \\\hdashline
24    & 0.00404(68)  &  0.0620(73) \\\hdashline
32    & 0.00152(15)  &  0.0365(61) \\
16+16 & 0.00164(33)  &  0.0396(71) \\\hdashline
48    & 0.00081(9)   &  0.0154(11) \\
24+24 & 0.00016(2)   &  0.0151(26) \\\hdashline
64    & 0.00063(6)   &  0.0163(22) \\
32+32 & 0.000052(2)  &  0.0047(5)  \\
\hline\hline
    \end{tabular}
\quad
    \begin{tabular}{c|llll}
\hline\hline
      & \multicolumn{2}{c}{Vector} & \multicolumn{2}{c}{Axial} \\
$N_z$ & $Q^2$(PCA) & $Q^2$(AE) & $Q^2$(PCA) & $Q^2$(AE) \\\hline
1     & 0.326(24)          & 0.160(16)        & 0.501(13)        & 0.250(11)  \\
2     & 0.1103(91)         & 0.0243(23)       & 0.2760(97)       & 0.0822(58) \\
3     & 0.0356(31)         & 0.00454(36)      & 0.1803(72)       & 0.0419(25) \\
4     & 0.00021(2)       & 0.00019(1)     & 0.1073(52)       & 0.0256(11) \\
\hline\hline
    \end{tabular}
    \caption{$Q^2$, defined in Eq.~\protect\eqref{eq:Q-value}, of the binary compression (BC) algorithm we propose with $N_q$ qubits (left) and classical approaches of the principal component analysis (PCA) and autoencoder (AE) with $N_z$ codes (right) for the vector and axial-vector data. The results with $N_q = N_{q1}+N_{q2}$ shows the compression with the boosting explained in Section~\protect\ref{sec:boosting}.
    Results are averaged over 10 independent sets, and the errors are calculated as the standard deviation of the mean. A smaller $Q^2$ indicates a better reconstruction. 
    }
    \label{tab:Q-simann}
\end{table}

\subsection{Experiments with D-Wave 2000Q}
To verify the usability of the existing quantum hardware for the proposed compression algorithm, we carry out the $\boldsymbol{a}^{(k)}$-optimization of Eq.~\eqref{eq:H_SC} using the D-Wave 2000Q quantum processor with the $\boldsymbol{\phi}$ obtained using the simulated annealing as described in Section~\ref{sec:exp_sim}. The major issue with the D-Wave quantum annealer is that the $h$ and $J$ parameters have poor precision when implemented in the D-Wave QPU, even though they are specified as double-precision floating-point numbers in the program, due to the integrated control errors~\citep{dwave_ice}. In the data compression problem, to minimize the effect of the fidelity loss in the final results, we restrict the maximum absolute value of the matrix elements of $\boldsymbol{\phi}$ by 1. It prevents a large maximum absolute value of $h$ and $J$, which introduces a large distortion of the small-value elements. Due to the limited D-Wave access time, we carry out the study only for one set of $N=200$ samples. For this study, we obtained solutions to this QUBO problem on the LANL 2000Q quantum D-Wave hardware that are drawn from 5000 reads using a series of 20 different chain strengths within the range (2.0, 3.0).  Results are taken from the lowest energy points among overall $5000 \times 20$ solutions of the 2000Q machine. The embedding procedure is repeated only for a new input but was kept unchanged as we changed the chain strength values.  In a control run, we find that even if one runs a new embedding each time a new chain strength changes, the final results do not differ from the method described above.  However, the later approach that requires a new embedding solution for each chain strength will require more pre-processing time.

Table~\ref{tab:dwave} shows the $Q^2$ values of the binary compression algorithm on D-Wave 2000Q in comparison with the simulated annealing optimizer. When $N_q \le 16$, D-Wave shows similar performance as the simulated annealing, but when $N_q > 16$, D-Wave shows worse performance than the simulated annealing. The reconstruction error, represented by $Q^2$, is decreased as $N_q$ is increased on the simulated annealing, but no significant decrease of the $Q^2$ is observed on the D-Wave for $N_q > 32$ compared to the results from $N_q=32$. As expected, constraining $\textrm{max}(|\boldsymbol{\phi}_{ij}|)=1$ improves the results on the D-Wave for $N_q \ge 32$, but the D-Wave results are still worse than the simulated annealing. 
Note that, due to the limited D-Wave access time, the results were obtained with a fixed $\boldsymbol{\phi}$ obtained using the simulated annealing. Hence, the results show a comparison of the optimization performance for a given problem. If $\boldsymbol{\phi}$ were obtained directly from the D-Wave quantum annealer, however, optimal constraints to meet the hardware limitations would have been imposed, naturally, and it might have resulted in a better compression performance than those of the $\textrm{max}(|\boldsymbol{\phi}_{ij}|)=1$ constraints.


\begin{table}[]
    \centering
    \begin{tabular}{c|cccc}
    \hline\hline
      & \multicolumn{4}{c}{$Q^2$ (Vector)}\\\hline
   &\multicolumn{2}{c}{Free $\boldsymbol{\phi}$} & \multicolumn{2}{c}{$\textrm{max}(|\boldsymbol{\phi}_{ij}|)=1$} \\\hline
 $N_q$ &   D-Wave   &  Sim.Ann.  &   D-Wave   &  Sim.Ann.   \\\hline
 8 &  0.104(11) &  0.104(11) &  0.099(10) &  0.099(10)  \\
16 & 0.0124(16) & 0.0120(16) & 0.0192(25) & 0.0197(27)  \\
32 & 0.0068(12) & 0.0014(02) & 0.0046(10) & 0.0033(06)  \\
48 & 0.0066(10) & 0.0007(01) & 0.0048(11) & 0.0015(03)  \\
60 & 0.0099(19) & 0.0007(01) & 0.0025(04) & 0.0006(01)  \\\hline\hline
    &\multicolumn{4}{c}{$Q^2$ (Axial)} \\\hline
   &\multicolumn{2}{c}{Free $\boldsymbol{\phi}$} & \multicolumn{2}{c}{$\textrm{max}(|\boldsymbol{\phi}_{ij}|)=1$} \\\hline
 $N_q$ &   D-Wave   &  Sim.Ann.  &   D-Wave   &  Sim.Ann.   \\\hline
 8 &  0.289(22) &  0.289(22) &  0.297(30) &  0.297(30) \\
16 & 0.1117(87) & 0.1101(87) &  0.129(20) &  0.135(20) \\
32 & 0.1113(86) & 0.0366(50) &  0.090(11) &  0.072(12) \\
48 &  0.092(15) & 0.0214(34) & 0.0751(71) & 0.0303(39) \\
60 & 0.0962(70) & 0.0175(16) & 0.0886(86) & 0.0221(20) \\\hline\hline
    \end{tabular}
    \caption{$Q^2$ values of the binary compression algorithm on D-Wave 2000Q quantum annealer (D-Wave) and the simulated annealing (Sim.Ann.) with ($\textrm{max}(|\boldsymbol{\phi}_{ij}|)=1$) and without (Free $\boldsymbol{\phi}$) the constraints on the elements of $\boldsymbol{\phi}$. Results are obtained from a set of $N=200$ vector and axial-vector data. Numbers in the parenthesis are the statistical error of the 200 samples estimated by the bootstrap method~\cite{Efron1992}.}
    \label{tab:dwave}
\end{table}

\subsection{Comparison of D-Wave 2000Q with Advantage Systems}
We benchmark the D-Wave Advantage system in comparison with the 2000Q using the $\boldsymbol{a}^{(k)}$-optimization problem in Eq.~\eqref{eq:H_SC}. For axial and vector data we compute the cumulative distribution function (CDF) of the normalized reconstruction error for systems of size $N_q=(32,60)$. To minimize possible biases due to a specific choice of embedding, we employ the heuristic solvers provided by Dwave to find an embedding for each configuration and proceed to collect at least 1500 samples (per configuration).  The chain strength during embedding was determined by the maximal coupling in absolute value, multiplied by a hyperparameter which we call chain strength multiple. The number of physical qubits, in practice, is many times higher than the logical qubits required, due to hardware connectivity.
There are cases where the physical qubits that are strongly coupled to behave as one logical qubit, return different values and we discard these samples, as non viable solutions, from our calculation of the distribution function. 
(For axial data, $N_q=32$: 150 qubits Advantage / 350 qubits 2000Q, $N_q=60$: 600 qubits Advantage / 1600 qubits 2000Q. For vector data, $N_q=32$: around 180 qubits Advantage / 380 qubits 2000Q, $N_q=60$:  600 qubits Advantage / 1400 qubits 2000Q). As the number of qubits increases the fraction of feasible samples decreases.  Also, the fraction of the CDF with small reconstruction error decreases. 
By trial and error, we find that setting chain strength multiple to a value greater than 1 reduces the number of viable solutions from the Advantage system, but it improves the results from 2000Q. For the samples collected for axial data, we set them to 0.8 and 1.6 respectively.

 \begin{figure}[ht]
    \centering
    \subfloat[$N_q=32$]{\includegraphics[width=0.48\textwidth]{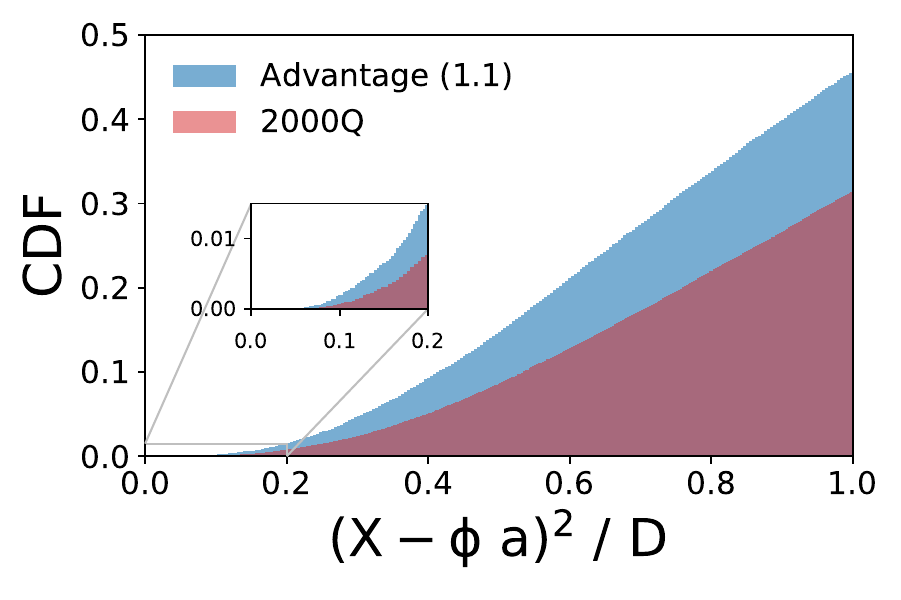}}
    \subfloat[$N_q=60$]{\includegraphics[width=0.48\textwidth]{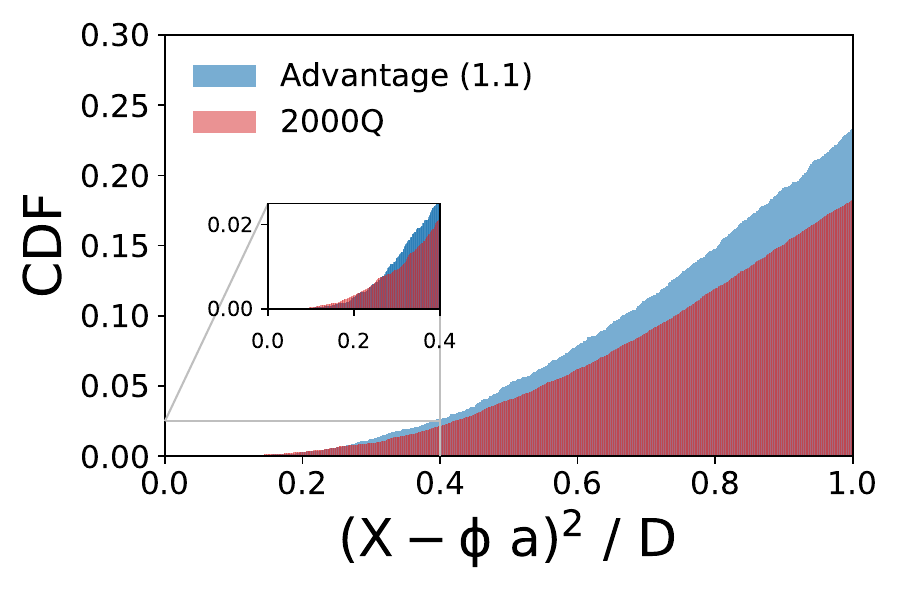}}
    \caption{Cumulative distribution function (CDF) of the normalized reconstruction error from all feasible samples obtained from the D-Wave 2000Q (red) and Advantage system (blue) for the axial-vector data. About 50\% and 38\% of the samples were feasible from D-Wave 2000Q and Advantage for $N_q=32$, respectively. For $N_q=60$ there were about 51\% and 18\%, respectively.}
    \label{fig:cdf_2000Q_vs_Adv_axial}
\end{figure}

\begin{figure}[ht]
    \centering
    \subfloat[$N_q=32$]{\includegraphics[width=0.48\textwidth]{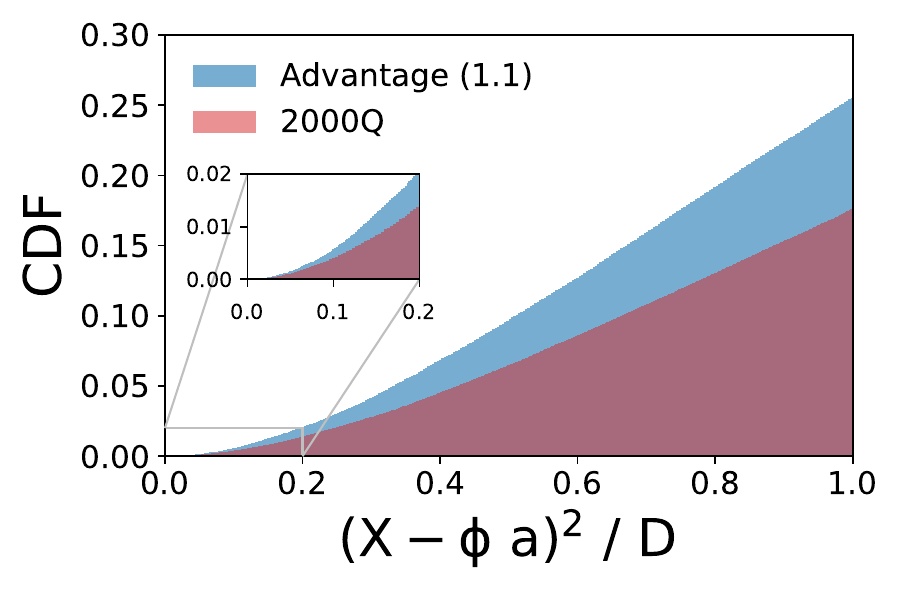}}
    \subfloat[$N_q=60$]{\includegraphics[width=0.48\textwidth]{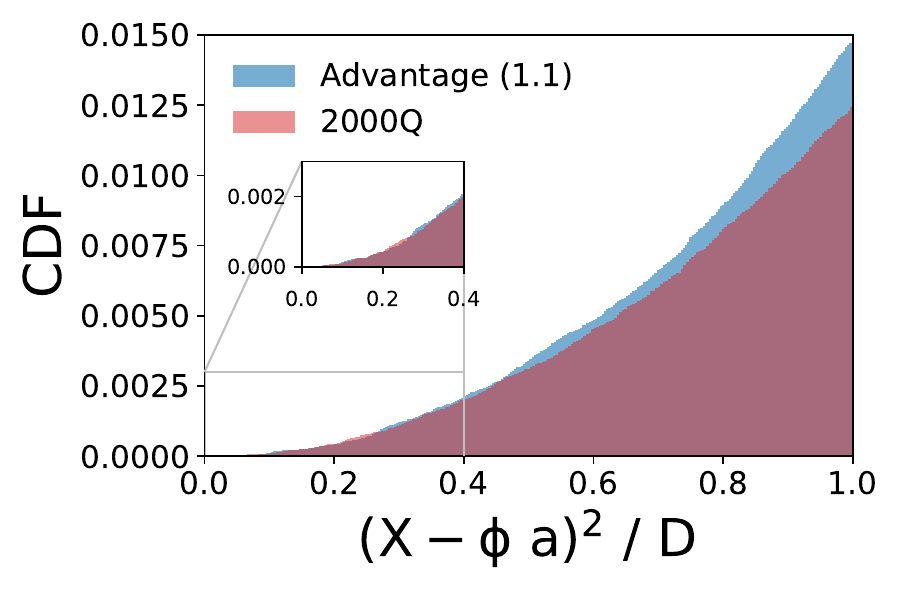}}
    \caption{Cumulative distribution function (CDF) of the normalized reconstruction error from all feasible samples obtained from the D-Wave 2000Q (red) and Advantage system (blue) for the vector data. About 95\% and 91\% of the samples were feasible from D-Wave 2000Q and Advantage for $N_q=32$, respectively. For $N_q=60$ there were about 63\% and 73\%, respectively.}
    \label{fig:cdf_2000Q_vs_Adv_vector}
\end{figure}

As can be seen from Figures~\ref{fig:cdf_2000Q_vs_Adv_axial} and ~\ref{fig:cdf_2000Q_vs_Adv_vector}, when $N_q=32$, both hardware perform rather well, and the new Advantage system has better statistics and overall higher quality of sub optimal solutions. In the case of $N_q=60$, the difference between the two hardware becomes less distinct and the CDF is peaked on solutions with high reconstruction error. As we did not apply boosting for these experiments, the quality degradation as the number of qubits increased can be ascribed to the connectivity of the hardware.
 
\section{Discussion}
In this paper, we presented a new lossy compression algorithm for statistical data based on the representation learning with binary variables. The algorithm finds a set of basis vectors, which is common for all data, and their binary coefficients ($N_q$) that precisely reconstruct each $D$-dimensional input vector. The algorithm provides data compression because the $N_q$-dimensional binary representation requires much smaller storage space than the original data of $D$-dimensional floating-point numbers. We also presented a bias correction procedure estimating the errors due to the inexact reconstruction of the lossy compression in Section~\ref{sec:BC}. The compression algorithm was applied to two lattice QCD datasets in Section~\ref{sec:expt}. With simulated annealing, the binary compression algorithm was able to achieve the same quality of reconstruction with 3.5 times smaller storage usage than the algorithms using neural-network autoencoder and PCA. The binary optimization carried out on D-Wave 2000Q for the compression problems showed promising results, but the performance was limited by the integrated control error of the D-Wave QPU, which introduces large uncertainties in the $h$ and $J$ parameters. The comparison of D-Wave 2000Q and Advantage systems showed that the Advantage is more efficient than the 2000Q in obtaining the low-energy solutions.

The proposed compression algorithm is a natural outlier detector because input data with large reconstruction errors can be marked anomalous~\citep{Adler:2015}. Using the proposed algorithm, furthermore, many operations that need to be performed on the floating point numbers $\mathbf{X}^{(k)}$ can be replaced by those on single-bit coefficients $\boldsymbol{a}^{(k)}$ with much smaller computational cost, because the relationship between $\mathbf{X}^{(k)}$ and $\boldsymbol{a}^{(k)}$ is linear ($\mathbf{X}^{(k)} \approx \boldsymbol{\phi} \boldsymbol{a}^{(k)}$), and the single-bit coefficients satisfy $\left(a_j^{(k)}\right)^n = a_j^{(k)}$ for any $n$, which simplifies power operations. Here are two examples of the operations in the compressed space:
\begin{itemize}
\item Sum of vectors
\begin{align}
  \sum_{k=1}^N \mathbf{X}^{(k)} & \approx \sum_{k=1}^N \boldsymbol{\phi} \boldsymbol{a}^{(k)}  = \boldsymbol{\phi} \left(\sum_{k=1}^N  \boldsymbol{a}^{(k)}\right)\,,
\end{align}
\item Sum of $l^2$-norm squares
\begin{align}
  \sum_{k=1}^N ||\mathbf{X}^{(k)}||^2 & \approx \sum_{k=1}^N \sum_{i=1}^D\left(\sum_{j=1}^{N_q} \phi_{ij} a_j^{(k)} \right)^2 \nonumber \\
  & = \sum_{i=1}^D \left[ \sum_{j=1}^{N_q} \phi_{ij}^2 \left( \sum_{k=1}^N a_j^{(k)} \right)
  + 2\sum_{l<m} \left(\sum_{k=1}^N a_l^{(k)} a_m^{(k)}\right) \phi_{il} \phi_{im} \right]\,.
\end{align}
\end{itemize}
The cost reduction is maximized when $D, N_q \ll N$, which is a typical case of many statistical datasets.

In this study, we presented only the results with the $\boldsymbol{\phi}$ calculated from the whole dataset. In general, however, $\boldsymbol{\phi}$ obtained from a smaller subset of the whole data provides a reasonably good compression performance. When using a $\boldsymbol{\phi}$ obtained from a subset data, some unseen data vectors could yield large reconstruction error. To control the error and maintain the quality of the compression, one needs to define a threshold and save the original data when the data gives a reconstruction error bigger than the threshold.

\section*{Acknowledgments}
The QUBO optimizations were carried out using the D-Wave 2000Q at Los Alamos National Laboratory (LANL) and the D-Wave's Leap Quantum Cloud Service. Simulation data used for the numerical experiment were generated using the computer facilities at (i)the National Energy Research Scientific Computing Center, a DOE Office of Science User Facility supported by the Office of Science of the U.S. Department of Energy under Contract No. DE-AC02-05CH11231; and, (ii) the Oak Ridge Leadership Computing Facility at the Oak Ridge National Laboratory, which is supported by the Office of Science of the U.S. Department of Energy under Contract No. DE-AC05-00OR22725; (iii) the USQCD Collaboration, which is funded by the Office of Science of the U.S. Department of Energy, (iv) Institutional Computing at Los Alamos National Laboratory. This work was supported by the U.S. Department of Energy, Office of Science, Office of High Energy Physics under Contract No. 89233218CNA000001. Lawrence Berkeley National Laboratory (LBNL) is operated by The Regents of the University of California (UC) for the U.S. Department of Energy (DOE) under Federal Prime Agreement DE-AC02-05CH11231. This material is based upon work supported by the U.S. Department of Energy, Office of Science, Office of Nuclear Physics, Quantum Horizons: QIS Research and Innovation for Nuclear Science under Award Number FWP-NQISCCAWL (CCC). ER acknowledges the NSF N3AS Physics Frontier Center, NSF Grant No. PHY-2020275, and the Heising-Simons Foundation (2017-228).

\appendix 
\section{Performance of Simulated Annealing Sampler}\label{sec:perf_sim}
In this section, we evaluate the quality of the D-Wave's simulated annealing sampler. Consider a problem finding $N_q$ binary coefficients $\boldsymbol{a}\in \{0,1\}^{N_q}$ of the positive powers of $r^{-1}<1$ that precisely reconstruct a uniform random number $z \in [0,1)$ by minimizing the following reconstruction error
\begin{align}
    E = \left\vert z -  \frac{1}{R} \sum_{n=1}^{N_q} a_n r^{-n}\right\vert \quad \textrm{ where } R \equiv \sum_{n=1}^{\infty} r^{-n}\,.
\label{eq:num_recon}
\end{align}
The optimization problem can be converted into QUBO form of Eq.~\eqref{eq:QUBO} by the taking the transformation given in Eq.~\eqref{eq:QUBO-transf} after replacing $\boldsymbol{\phi}$ and $\mathbf{X}$ with the vector of $\{r^{-n} / R\}$ and $z$, respectably,
\begin{align*}
  \boldsymbol{\phi} \rightarrow 
  \begin{bmatrix}
    r^{-1} / R \\
    r^{-2} / R \\
    r^{-3} / R \\
    \vdots \\
    r^{-N_q} / R \\
  \end{bmatrix},
  \qquad
  \mathbf{X} \rightarrow z\,.
\end{align*}
We solve the QUBO problem using D-Wave's simulated annealing sampler on a classical computer by taking the minimum energy solution from the three different choices of $\emph{num\_reads} = 10, 30$, and $150$, keeping other sampler parameters set to default, for $10^5$ random numbers of $r$ and calculate the average value of the reconstruction error $E$. The study is done at two different values of $r=2$ and $1.5$. When $r=2$, it becomes a simple decimal to binary conversion problem, whose optimal solution is known. The expected value of the average reconstruction error for an ideal QUBO solver for $N_q \gg 1$ is $2^{-(N_q+2)}$. For $r=1.5$, we calculate the empirical average reconstruction error of an ideal QUBO solver by fitting the average values of the reconstruction errors obtained using the exact solver implemented in the D-Wave Ocean library~\cite{dwave_ocean}, which finds the minimum energy solution by comparing the energies of all possible solutions. The two free parameters of the fitting functional form $a^{N_q+b}$ are determined to be $a=0.514(2)$ and $b=1.01(6)$ from the 7 data points at $N_q = 8,10,12,\ldots,20$ with the $\chi^2/\textrm{dof}=1.26$.

\begin{figure}
    \centering
    \includegraphics[width=0.47\textwidth]{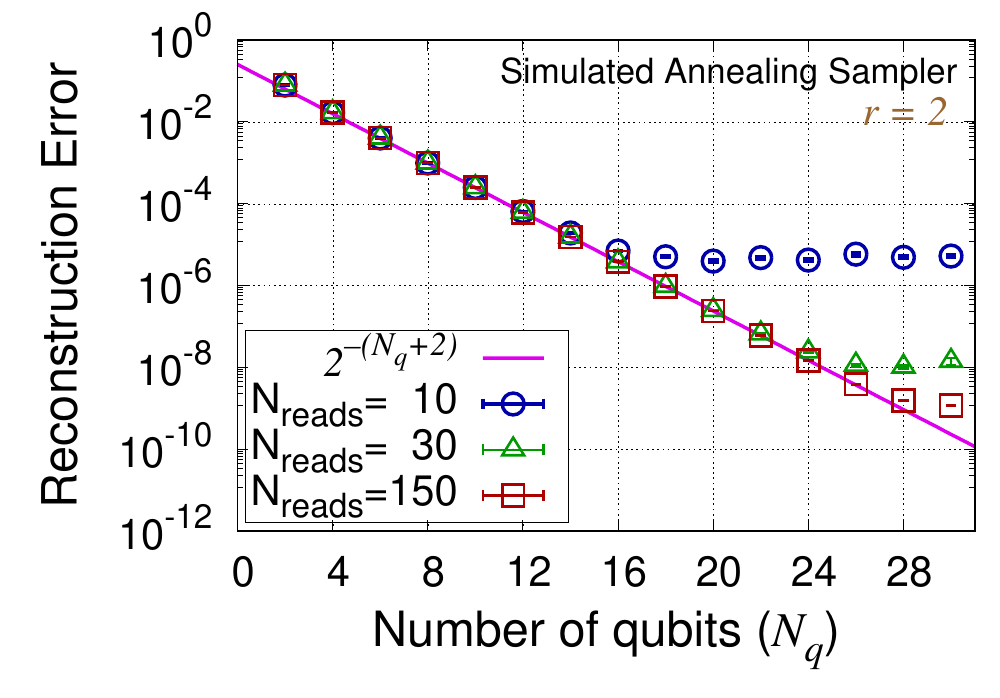}\qquad
    \includegraphics[width=0.47\textwidth]{{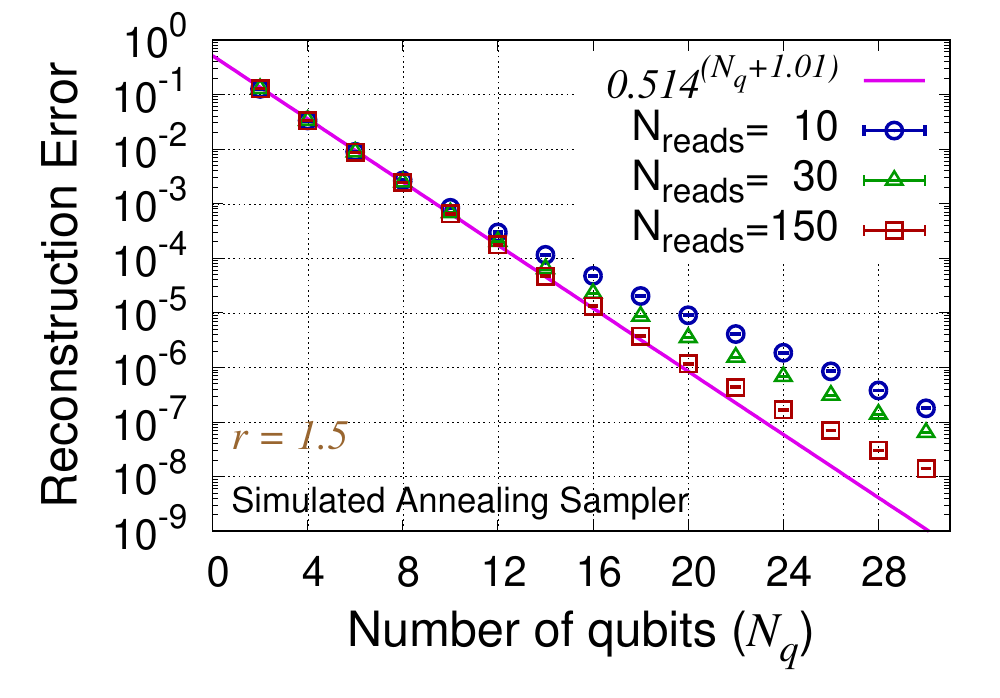}}
    \caption{Average reconstruction error defined in Eq.~\protect\eqref{eq:num_recon} for different number of reads (10, 30, and 150) of the simulated annealing sampler for $r=2$ (left) and $r=1.5$ (right). The expected average reconstruction error of the ideal QUBO solver is plotted as a magenta line.
    }
    \label{fig:pwr_scaling}
\end{figure}

Figure~\ref{fig:pwr_scaling} shows that a larger number of reads makes the reconstruction error smaller, the problems with the larger number of qubits require the larger number of reads to make the solution close to the exact solution. Results show that the simulated annealing sampler with {\emph{num\_reads}=150} gives the solution close to the ground-energy up to $N_q \approx 20$, but it may depend on the problem, as demonstrated by the difference between the $r=1.5$ and $r=2$ cases.

\bibliographystyle{apsrev4-1}
\bibliography{ref}

\end{document}